\documentclass[twocolumn,showpacs,preprintnumbers,amsmath,amssymb]{revtex4}

\usepackage{graphicx}
\usepackage{dcolumn}
\usepackage{bm}
\usepackage{graphics}
\usepackage{amsmath}
\usepackage{amssymb}
\usepackage{amscd}
\usepackage{afterpage}
\usepackage{float,times}
\usepackage{subfigure}
\usepackage{rotating}
\usepackage{multirow}
\usepackage{fancyheadings}
\usepackage{epsfig}
\usepackage{theorem}
\usepackage{moreverb}
\usepackage{euscript}
\usepackage{psfrag}

\begin{document}

\title{Evaluating The Two Loop Diagram Responsible For \\Neutrino Mass In Babu's Model}

\author{K. L. McDonald}\email{k.mcdonald@physics.unimelb.edu.au}

\author{B. H. J. McKellar}%
 \email{b.mckellar@physics.unimelb.edu.au}
\affiliation{%
School of Physics, Research Centre for High Energy Physics, The
University of Melbourne, Victoria, 3010, Australia\\
}%

\date{\today}

\begin{abstract}
Babu studied the neutrino spectrum obtained when
one adds a charged singlet and a doubly charged
singlet to the standard model particle spectrum. It was found that the neutrinos acquire a mass
matrix at the two-loop level which contains one massless
eigenstate. The mass matrix of Babu's model depends on an
integral over the undetermined loop momenta. We present the exact calculation of this integral.
\end{abstract}

\maketitle

\section{\label{sec:intro}INTRODUCTION\protect\\}

Zee observed that the addition of a charged singlet and additional
Higgs doublets to the Standard
Model particle spectrum resulted in radiatively generated neutrino
mass at the one-loop level~\cite{zee}. It was consequently noted that if only one
doublet couples to the leptons, the mass matrix takes a simple form and
produces one mass eigenstate much lighter than the other
two~\cite{wolf_zee}. Babu studied the neutrino spectrum obtained when
one retains Zee's charged singlet and adds a doubly charged
singlet~\cite{babu}. It was found that the neutrinos develop a mass
matrix at the two-loop level, which contains one massless eigenstate,
to lowest order. The mass matrix depends on an
integral over the undetermined loop momenta of the two-loop
diagram responsible for the neutrino mass, for which Babu gave an
approximate form. We present the exact
calculation of this integral. The analytic evaluation of related integrals,
occurring when doublet neutrinos acquire mass at the two-loop level via
the exchange of $W^{\pm}$ bosons in
the presence of singlet neutrinos~\cite{petkov_toshev, babu_ma}, can be found in~\cite{gracey}.

In Sec. \ref{sec:outline_babu} we briefly review Babu's model for
the generation of neutrino
mass. Sec. \ref{sec:analytic_evaluation_of_i_cd} contains the analytic
evaluation of the integral which
sets the scale for the massive neutrinos in Babu's model. The leading
terms of this integral are presented in Sec. \ref{sec:leading_terms}
for the relevant hierarchies of mass parameters involved. 
\section{Babu's Model\label{sec:outline_babu}}
\begin{figure}[b]
\begin{center}

\includegraphics{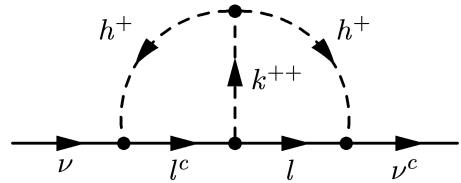}
\caption{Two-loop diagram responsible for neutrino mass in the Babu
  model.}
\label{fig:babus_two_loop_diagram}
\end{center}
\end{figure}

Babu's model~\cite{babu} includes two $SU(2)_{L}$ singlet Higgs fields; a singly charged field
\mbox{$h^{+}$} and a doubly charged field \mbox{$k^{++}$}. The
addition of these singlets gives rise to the Yukawa couplings:
\begin{eqnarray}
\mathcal{L_{Y}}=f_{ab}\overline{(\Psi_{aL})^{C}}\Psi_{bL}h^{+}+h_{ab}\overline{(l_{aR})^{C}}l_{bR}k^{++}+h.c.
\end{eqnarray}
where \mbox{$f_{ab}=-f_{ba}$} and \mbox{$h_{ab}=h_{ba}$}.
Gauge invariance precludes the singlet Higgs fields from coupling to
the quarks. The Higgs potential contains the terms:
\begin{equation}
V(\phi, h^{+},
k^{++})=\mu(h^{-}h^{-}k^{++}+h^{+}h^{+}k^{--}) + ......
\end{equation}
which violate lepton number by two
units and give rise to neutrino Majorana mass contributions at the two-loop
level (see Figure~\ref{fig:babus_two_loop_diagram}).
 The neutrino masses are calculable and to lowest order the mass matrix takes the form:
\begin{eqnarray}
M_{ab}=8\mu f_{ac}\tilde{h}_{cd}m_{c}m_{d}I_{cd}(f^{\dag})_{db}, 
\end{eqnarray}
where \mbox{$\tilde{h}_{ab}=\eta h_{ab}$} with \mbox{$\eta =1$} for
\mbox{$a=b$} and \mbox{$\eta =2$} for
\mbox{$a\ne b$}. $m_{c,d}$ are the charged lepton masses and:
\begin{eqnarray}
I_{cd}&=&\int \frac{d^{4}p}{(2\pi)^{4}}\int
\frac{d^{4}q}{(2\pi)^{4}}\frac{1}{(p^{2}-m_{h}^{2})}\frac{1}{(p^{2}-m_{c}^{2})}\frac{1}{(q^{2}-m_{h}^{2})}\nonumber\\
& &\times\frac{1}{(q^{2}-m_{d}^{2})}\frac{1}{(p-q)^{2}-m_{k}^{2}}.
\end{eqnarray}
If one defines:
\begin{eqnarray*}
K_{cd}=8\mu \tilde{h}_{cd}m_{c}m_{d}I_{cd},
\end{eqnarray*}
where no summation is implied by the repeated indices, the mass matrix may be written as
$M_{ab}=(fKf^{\dagger})_{ab}$. Thus $\mathrm{Det}M=|\mathrm{Det}
 \ f|^{2}\mathrm{Det} K=0$ for an odd number of generations (due to the anti-symmetry of $f$) and to
lowest order the neutrino spectrum contains one massless state.
In what follows, the exact calculation of the integral $I_{cd}$ is
performed and the asymptotic behaviour for the cases \mbox{$m_{k}\gg
m_{h}\gg m_{c}, m_{d}$} and \mbox{$m_{h}\gg
m_{k}\gg m_{c}, m_{d}$} is presented. Note that the hierarchy \mbox{$m_{h},
m_{k}\gg m_{c}, m_{d}$} is a phenomenological constraint, whilst the
relative size of $m_{k}$ and $m_{h}$ is not predetermined.

\section{\label{sec:analytic_evaluation_of_i_cd}Evaluation Of $\mathbf{I_{cd}}$}
After performing a Wick rotation and letting $q\rightarrow -q$,
the integral $I_{cd}$ may be written
as:
\begin{eqnarray}
I_{cd}&=&\int \frac{d^{4}p}{(2\pi)^{4}}\int
\frac{d^{4}q}{(2\pi)^{4}}\frac{1}{(p^{2}+m_{h}^{2})}\frac{1}{(p^{2}
+m_{c}^{2})}\frac{1}{(q^{2}+m_{h}^{2})}\nonumber\\
& &\times\frac{1}{(q^{2}+m_{d}^{2})}\frac{1}{(p+q)^{2}+m_{k}^{2}}
\end{eqnarray}
where the momenta $p$ and $q$ are now Euclidean four-vectors.
We adopt the notation of~\cite{vanderbij_veltman} and define:
\begin{eqnarray}
& &(M_{11},M_{12}...M_{1n_{1}}|M_{21},M_{22}...m_{2n_{2}}|M_{31},M_{32}...M_{3n_{3}})\nonumber\\
&=&\int d^{n}p\int
d^{n}q\prod_{i=1}^{n_{1}}\prod_{j=1}^{n_{2}}\prod_{k=1}^{n_{3}}\frac{1}{(p^{2}+M_{1i}^{2})}\times\nonumber\\
& &\mkern80mu\frac{1}{(q^{2}+M_{2j}^{2})}\frac{1}{((p+q)^{2}+M_{3k}^{2})},\label{vanderbij_notation}
\end{eqnarray} 
so that $I_{cd}=\frac{1}{(2\pi)^{8}}(m_{h},m_{c}|m_{h},m_{d}|m_{k})$ when $n=4$, where $n$
is the space-time dimensionality. Note that all momenta in
(\ref{vanderbij_notation}) are Euclidean whilst the definition used
in~\cite{vanderbij_veltman} contains Minkowski vectors. It is possible to
express $I_{cd}$ as a linear combination of integrals with less than five propagators.
One may use partial fractions to obtain relations like:
\begin{eqnarray*}
(m,m_{0}|m_{1}|m_{2})=\frac{1}{m^{2}-m_{0}^{2}}\left\{(m_{0}|m_{1}|m_{2})-(m|m_{1}|m_{2})\right\},
\end{eqnarray*}
which may be applied to $I_{cd}$ to give:
\begin{eqnarray}
& &I_{cd}=\frac{1}{(2\pi)^{8}}\frac{1}{(m_{h}^{2}-m_{c}^{2})}\frac{1}{(m_{h}^{2}-m_{d}^{2})}\left\{(m_{c}|m_{d}|m_{k})-\right.\nonumber\\ 
& &\left.(m_{h}|m_{d}|m_{k})-(m_{a}|m_{h}|m_{k})+(m_{h}|m_{h}|m_{k})\right\}.
\end{eqnarray}
At this point one may proceed to evaluate the integral
$(m_{0}|m_{1}|m_{2})$ to determine $I_{cd}$ but it is
preferable to express $I_{cd}$ as a combination of integrals of the
form $(2m_{0}|m_{1}|m_{2})$ (where $(2m_{0}|m_{1}|m_{2})$ is shorthand for
$(m_{0},m_{0}|m_{1}|m_{2})$). If one evaluates integrals of the form
$(m_{0}|m_{1}|m_{2})$ by introducing Feynman parameters, some of the
ultraviolet divergences are transferred from the radial
integrals in momentum space to the Feynman parameter integrals. It is
easier to handle the divergences when evaluating
$(2m_{0}|m_{1}|m_{2})$, as they may be completely contained in
the momentum space integrals. Integrals of the type
$(m_{0}|m_{1}|m_{2})$ have been considered in~\cite{lahanes_tamvakis_vayonakis,
  broadhurst} and single integral representations have been
obtained~\cite{davydychev_tausk, ford_jones, ford_jack_jones}.

To express $(m_{0}|m_{1}|m_{2})$ in terms of integrals of the form
$(2m_{0}|m_{1}|m_{2})$ one may use the partial $p$ operation of 't~Hooft~\cite{'thooft_veltman}. Essentially one inserts the identity
expression:
\begin{eqnarray}
\label{partial_p}
1=\frac{1}{2n}\sum_{i=1}^{n}\left\{\frac{\partial p_{i}}{\partial
    p_{i}} +\frac{\partial q_{i}}{\partial
    q_{i}}\right\}
\end{eqnarray}
into the integrand of $(m_{0}|m_{1}|m_{2})$, performs integration by
parts and rearranges the resulting expressions to obtain the
relationship:
\begin{eqnarray}
(m_{0}|m_{1}|m_{2})=\frac{1}{3-n}\times\left\{m_{0}^{2}(2m_{0}|m_{1}|m_{2})\right.\nonumber\\
\left.+ m_{1}^{2}(2m_{1}|m_{0}|m_{2}) + m_{2}^{2}(2m_{2}|m_{0}|m_{1})\right\}.\label{mmm_to_mmmm}
\end{eqnarray} 
We note that equation (\ref{partial_p}) is defined for integral $n$, but
the resulting relationship may be analytically continued to
non-integral $n$. Using equation (\ref{mmm_to_mmmm}) on $I_{cd}$ gives:
\begin{widetext}
\begin{eqnarray}
& &I_{cd}=\frac{1}{(2\pi)^{8}}\frac{1}{(m_{h}^{2}-m_{c}^{2})}\frac{1}{(m_{h}^{2}-m_{d}^{2})}\times\left\{\mkern3mum_{c}^{2}\left[(2m_{c}|m_{h}|m_{k})-(2m_{c}|m_{d}|m_{k})\right]+m_{d}^{2}\left[(2m_{d}|m_{h}|m_{k})-(2m_{d}|m_{c}|m_{k})\right]\right.\nonumber\\
& &\mkern310mu +\ 
m_{k}^{2}\left[(2m_{k}|m_{h}|m_{d})+(2m_{k}|m_{c}|m_{h})-(2m_{k}|m_{c}|m_{d})
-(2m_{k}|m_{h}|m_{h}) \right]\nonumber\\
& &\left.\mkern310mu +\  m_{h}^{2}\left[(2m_{h}|m_{d}|m_{k})+(2m_{h}|m_{c}|m_{k})
  -2(2m_{h}|m_{h}|m_{k})\right]\mkern4mu \right\}.\label{i_cd_vanderbij_expressions}
\end{eqnarray}
Thus evaluation of the generic integral $(2m|m_{1}|m_{2})$ allows
one to determine $I_{cd}$. We obtain: 
\begin{eqnarray}
(2m|m_{1}|m_{2})&=&\frac{\pi^{4}(\pi
m^{2})^{n-4}\Gamma(2-\frac{1}{2}n)}{\Gamma(3-\frac{1}{2}n)} \int_{0}^{1}dx\int_{0}^{1}dy(x(1-x))^{n/2
-2}y(1-y)^{2-n/2}\times\nonumber\\
& & \left[ \Gamma (5-n)\frac{\mu^{2}}{(y + \mu^{2}(1-y))^{5-n}}
+\frac{1}{2}n\Gamma (4-n)\frac{1}{(y+\mu^{2}(1-y))^{4-n}}\right],\nonumber
\end{eqnarray} 
\end{widetext}
where
\begin{displaymath}
\mu^{2}\equiv \frac{ax+b(1-x)}{x(1-x)},
\mkern25mu  a\equiv \frac{m^{2}_{1}}{m^{2}}, \mkern25mu b\equiv \frac{m^{2}_{2}}{m^{2}}.
\end{displaymath}
This result is in agreement with the result for
$\mathcal{G}(m,m_{1},m_{2};0)$ in~\cite{ghinculov_vanderbij} and differs by an overall minus sign to that
obtained in~\cite{vanderbij_veltman} due to our different definition
of $(2m|m_{1}|m_{2})$ in terms of Euclidean momenta. Letting
$n=4+\epsilon$ and expanding for
small $\epsilon$ gives:
 \begin{eqnarray}
& &(2m|m_{1}|m_{2})\nonumber\\
&=& -\pi^{4}\left[ \frac{-2}{\epsilon^{2}} +
\frac{1}{\epsilon}(1-2\gamma_{E}-2\log(\pi m^{2}))\right]\nonumber \\
& & - \pi^{4}\left[ -\frac{1}{2}
-\frac{1}{12}\pi^{2}-\gamma^{2}_{E}+(1-2\gamma_{E})\log(\pi m^{2})\right.\nonumber\\
& & \left. \mkern45mu -\log ^{2}(\pi m^{2}) - f(a,b)\right] +
O(\epsilon ) \label{mm_mone_mtwo}
\end{eqnarray}
where the function $f(a,b)$ is given by:
\begin{displaymath}
f(a,b)=\int_{0}^{1}dx\left(\mathrm{Li_{2}}(1-\mu^{2}) - \frac{\mu^{2}\log \mu^{2}}{1-\mu^{2}}\right)
\end{displaymath}
and the dilogarithm function is defined as:
\begin{displaymath}
\mathrm{Li_{2}}(x)=-\int_{0}^{x}\frac{\log(1-y)}{y}dy.
\end{displaymath}
Evaluation of $f(a,b)$ gives:
\begin{widetext}
\begin{eqnarray}
& &f(a,b)= -\frac{1}{2}\log a\log b
-\frac{1}{2}\left(\frac{a+b-1}{\sqrt{\mkern15mu}}\right)\left\{\mathrm{Li_{2}}\left(\frac{-x_{2}}{y_{1}}
  \right)+
  \mathrm{Li_{2}}\left(\frac{-y_{2}}{x_{1}}\right)-\mathrm{Li_{2}}\left(\frac{-x_{1}}{y_{2}}\right)-\mathrm{Li_{2}}\left(\frac{-y_{1}}{x_{2}}\right) \right.\nonumber\\
& &\label{my_form_of_f}\mkern330mu \left. + \mathrm{Li_{2}}\left(\frac{b-a}{x_{2}}\right)+\mathrm{Li_{2}}\left(\frac{a-b}{y_{2}}\right)-
\mathrm{Li_{2}}\left(\frac{b-a}{x_{1}}\right)-\mathrm{Li_{2}}\left(\frac{a-b}{y_{1}}\right)\right\}. 
\end{eqnarray}
\end{widetext}
We have introduced:
\begin{eqnarray*}
x_{1}=\frac{1}{2}(1+b-a + \sqrt{\mkern15mu}),\mkern25mu
x_{2}=\frac{1}{2}(1+b-a - \sqrt{\mkern15mu}),\nonumber\\
y_{1}=\frac{1}{2}(1+a-b + \sqrt{\mkern15mu}),\mkern25mu y_{2}=\frac{1}{2}(1+a-b - \sqrt{\mkern15mu}),
\end{eqnarray*}
and:
\begin{eqnarray*}
\sqrt{\mkern15mu} =(1-2(a+b)+(a-b)^{2})^{1/2},
\end{eqnarray*} 
where under
\mbox{$a\leftrightarrow b$} we have \mbox{$x_{i} \leftrightarrow
  y_{i}$} for \mbox{$i=1,2$}. The expression (\ref{my_form_of_f}) can
be shown to be equivalent to:
\begin{eqnarray}
& &f(a,b)= -\frac{1}{2}\log a\log b -\left(\frac{a+b-1}{\sqrt{\mkern15mu}} -\frac{1}{2} \right)\times\nonumber\\
& &\left\{
\mathrm{Li_{2}}\left(\frac{-x_{2}}{y_{1}}\right)+\mathrm{Li_{2}}\left(\frac{-y_{2}}{x_{1}}\right)
+ \frac{1}{4}\log^{2}\frac{x_{2}}{y_{1}} +\frac{1}{4}\log^{2}\frac{y_{2}}{x_{1}}\right.
\nonumber\\
& & \left. \mkern65mu + \frac{1}{4}\log^{2}\frac{x_{1}}{y_{1}}
  -\frac{1}{4}\log^{2}\frac{x_{2}}{y_{2}} + \mathrm{Li_{2}}(1)\right\}\label{f_ab}
\end{eqnarray}
which is the symmetrized version obtained
in~\cite{vanderbij_veltman}. For numerical evaluation via mathematica
etc, it may be more convenient to use the form
(\ref{my_form_of_f}). If one uses (\ref{f_ab}) for arbitrary masses it
is possible for the logarithms to go negative, thus producing a
non-zero imaginary component for $f(a,b)$. By judicious logarithmic
branch choice one can ensure that the imaginary components of $f(a,b)$ cancel, whilst the
real part of $f(a,b)$ is independent of the branch choices. This real
part is always in agreement with the form (\ref{my_form_of_f}) of
$f(a,b)$, but in the latter the imaginary parts cancel when using
the principal branch for all dilogarithms. The result for
$I_{cd}$ will ultimately be a combination of functions of the form
$f(a,b)$. As $I_{cd}$ is evaluated in the neutrino rest frame, it is not
possible to cut the two-loop diagram it represents and obtain an energy-
and momentum-conserving sub-graph (ie the neutrino at rest doesn't
have enough energy for the internal particles to be real and on the
mass shell). Thus the amplitude for the graph must be
real~\cite{coleman_norton}, forcing the function $f(a,b)$ to be real.
Equation (\ref{mm_mone_mtwo}) together with either (\ref{f_ab}) or
(\ref{my_form_of_f}) gives the final result for
$(2m|m_{1}|m_{2})$.

 $I_{cd}$ is now obtained by
substituting the result
for $(2m|m_{1}|m_{2})$ into the expression
(\ref{i_cd_vanderbij_expressions}). Inspection of
(\ref{i_cd_vanderbij_expressions}) shows that we may use
$(2m|m_{1}|m_{2})\rightarrow\pi^{4} f(a,b)$ when evaluating $I_{cd}$, as the
constants and logarithms with massive arguments occurring in
$(2m|m_{1}|m_{2})$ cancel amongst the terms with a given mass
coefficient.

\section{\label{sec:leading_terms}Dominant Behaviour of $\mathbf{I_{cd}}$}
We now find the asymptotic behaviour for
the cases \mbox{$m_{k}\gg
m_{h}$} and \mbox{$m_{h}\gg
m_{k}$}. The dominant terms are found by using the
expansions for $(2m|m_{1}|m_{2})$ presented
in~\cite{vanderbij_veltman}. Utilising the symmetry properties of
$(2m|m_{1}|m_{2})$ and the structure of the expression
(\ref{i_cd_vanderbij_expressions}) for $I_{cd}$ allows one to obtain
the leading terms relatively easily.
\subsubsection{$\mathbf{m_{k}\gg m_{h}.}$}
In the expression for $I_{cd}$, equation
(\ref{i_cd_vanderbij_expressions}), we may safely neglect the terms with lepton mass
coefficients. The terms with coefficient $m_{k}^{2}$ have the form
$(2m_{k}|m_{x}|m_{y})$ and the expansion for $(2m|M_{1}|M_{2})$, where
$m\gg M_{1}, M_{2}$, given in~\cite{vanderbij_veltman}, may be
used. All constants and logarithms of $m^{2}$ may be ignored as they
cancel amongst the various terms. The expansion is given in terms of
$a=(M_{1}/m)^{2}$ and $b=(M_{2}/m)^{2}$. By noting the form of
(\ref{i_cd_vanderbij_expressions}) and that
$(2m|M_{1}|M_{2})=(2m|M_{2}|M_{1})$, it is seen that
in the expansion of a term $(2m_{k}|m_{x}|m_{y})$, any term depending
only on $a$ or only on $b$ will occur in the expansion of another term
with a relative minus sign. So one need only retain terms containing
both $a$ and $b$. Any terms which contain a lepton mass will be
suppressed, meaning the term $(2m_{k}|m_{h}|m_{h})$
dominates. Defining $h=(m_{h}/m_{k})^{2}$ and using $(2m|M_{1}|M_{2})$
in~\cite{vanderbij_veltman}, keeping only the terms containing both $a$
and $b$, gives:
\begin{eqnarray}
& &(2m_{k}|m_{h}|m_{h})=-\pi^{4}\left\{h^{2} +6h^{3}-2\log (h)\left\{h^{2}
  + 5h^{3}\right\}\right. \nonumber\\
& &\mkern100mu \left. -[\log^{2}(h) +
  \frac{\pi^{2}}{3}]\left\{h^{2} +4h^{3}\right\} +....\right\}.\label{khh_large_k}
\end{eqnarray}
All the terms with a coefficient $m_{h}^{2}$ in $I_{cd}$ have the form
$(2m_{h}|m_{x}|m_{k})$ and will be expanded in terms of $h$ and
$(m_{x}/m_{k})^{2}$. The form of (\ref{i_cd_vanderbij_expressions}) shows that
any term in an expansion depending only on $h$ (not including the cross
terms when $m_{x}=m_{h}$) will occur in another
expansion with a relative minus sign. If $m_{x}$ is a lepton mass,
the leading contributions from $(2m_{h}|m_{x}|m_{k})$ will contain
factors $(m_{x}/m_{k})^{2}$ and consequently be suppressed. Thus to
leading order:
\begin{eqnarray}
& &m_{h}^{2}\left\{(2m_{h}|m_{d}|m_{k})+(2m_{h}|m_{c}|m_{k})
  -2(2m_{h}|m_{h}|m_{k})\right\}\nonumber\\
& &\mkern160mu \approx
-2m_{h}^{2}(2m_{h}|m_{h}|m_{k}).
\end{eqnarray}
Using the expansion of $(2M_{1}|M_{2}|m)$ for $m\gg M_{1},M_{2}$
from~\cite{vanderbij_veltman} and noting the above gives:
\begin{eqnarray}
& &(2m_{h}|m_{h}|m_{k})=-\pi^{4}\left\{-h -\frac{21}{4}h^{2}  +
  \frac{\pi^{2}}{3}(h+2h^{2}) \right. \nonumber\\
& & \left.+\log^{2} (h)\left\{h + 3h^{2}\right\} +\log (h)\left\{h
  + \frac{13}{2}h^{2}\right\}+..\right\}\label{hhk_large_k}
\end{eqnarray}
as the leading terms which contribute to $I_{cd}$. Combining the
expressions for $(2m_{h}|m_{h}|m_{k})$ and $(2m_{k}|m_{h}|m_{h})$ and
noting the appropriate coefficients from equation
(\ref{i_cd_vanderbij_expressions}) gives:
\begin{eqnarray}
I_{cd}\simeq
\frac{1}{(4\pi)^{4}}\frac{1}{m_{k}^{2}}\left\{\log^{2}\frac{m_{h}^{2}}{m_{k}^{2}}
  +\frac{\pi^{2}}{3}-1\right\} +
O\left(\frac{1}{m_{k}^{4}}\right)\label{i_cd_large_k},
\end{eqnarray}
where the lepton masses have been neglected relative to the scalar
masses in the factors
$(m_{h}^{2}-m_{c}^{2})^{-1}(m_{h}^{2}-m_{d}^{2})^{-1}$. Babu's
approximate form reproduces the $\log^{2}$ term of
(\ref{i_cd_large_k}) in the $h\rightarrow 0$ limit. It is necessary to
retain both $(2m_{h}|m_{h}|m_{k})$ and $(2m_{k}|m_{h}|m_{h})$ to
obtain the leading terms of (\ref{i_cd_large_k}). The leading terms of $I_{cd}$ may be obtained
by setting \mbox{$m_{c}=m_{d}=0$} but one may not take
$m_{h}=0$ when $m_{k}\gg m_{h}$ as $I_{cd}$ develops an infra-red
singularity. This singularity is seen to manifest itself in the leading terms of
equation (\ref{i_cd_large_k}) as a logarithmic singularity in the
limit \mbox{$m_{h}\rightarrow 0$}.  
\subsubsection{$\mathbf{m_{h}\gg m_{k}.}$}
The terms with lepton mass coefficients may again be neglected to obtain
an expansion in terms of $k=(m_{k}^{2}/m_{h}^{2})$. The
presence of the factors
\mbox{$(m_{h}^{2}-m_{c}^{2})^{-1}(m_{h}^{2}-m_{d}^{2})^{-1}\approx
m_{h}^{-4}$} in (\ref{i_cd_vanderbij_expressions}) means that the terms
with coefficient $m_{k}^{2}$ will be suppressed relative to those with
coefficient $m_{h}^{2}$. The expansion of $(2m|m|M)$ for $m> M$
in~\cite{vanderbij_veltman} may be used for $(2m_{h}|m_{h}|m_{k})$
where only terms originating in the $f(a,b)$ portion, as defined in
(\ref{mm_mone_mtwo}) need be retained. The result is:
\begin{eqnarray}
(2m_{h}|m_{h}|m_{k})&=&-\pi^{4}\left\{k
  +\frac{5}{36}k^{2}+ \right.\nonumber\\
& &\left.\log (k)\left\{-\frac{1}{2}k -\frac{1}{12}k^{2}\right\}+..\right\}\label{hhk_large_h}.
\end{eqnarray}
For $m_{h}\gg m_{k}$, contributions from the terms
$(2m_{h}|m_{x}|m_{k})$, where $m_{x}$ is a lepton mass, must be
included. The leading terms are those which do not contain
factors $(m_{x}^{2}/m_{h}^{2})$. The expansion for $(2m|M_{1}|M_{2})$
with $m >M_{1},M_{2}$ in~\cite{vanderbij_veltman} is again employed.
Retaining only those contributions which stem from $f(a,b)$ gives:
\begin{eqnarray}
(2m_{h}|m_{x}|m_{k})&=&-\pi^{4}\left\{-\frac{\pi^{2}}{6} + k
  +\frac{1}{4}k^{2} +\right. \nonumber\\
& &\left.\log (k)\left\{-k -\frac{1}{2}k^{2}\right\}+..\right\}\label{hxk_large_h},
\end{eqnarray} 
with this expression holding for $m_{x}=m_{c}$ and
$m_{x}=m_{d}$. Using (\ref{hhk_large_h}) and (\ref{hxk_large_h})
and neglecting terms with
coefficients $m_{c}^{2}, m_{d}^{2}$ and $m_{k}^{2}$ in (\ref{i_cd_vanderbij_expressions}), gives the result:
\begin{eqnarray}
I_{cd}\simeq\frac{1}{2^{7}\pi^{4}}\frac{1}{m_{h}^{2}}\left\{\frac{\pi^{2}}{6}
  +\frac{1}{2}\frac{m_{k}^{2}}{m_{h}^{2}}\log\frac{m_{k}^{2}}{m_{h}^{2}}\right\}+O\left(\frac{1}{m_{h}^{4}}\right).
\end{eqnarray}
Thus for large $m_{h}^{2}$ the leading term is:
\begin{eqnarray}
I_{cd}\approx\frac{1}{(4\pi)^{4}}\frac{2}{m_{h}^{2}}\frac{\pi^{2}}{6}.\label{leading_i_large_h}
\end{eqnarray}
The $m_{h}^{-2}$ factor is expected on dimensional grounds and in this
limit the coefficient
of $m_{h}^{-2}$ contains no logarithmic singularities. We see that the dominant terms for $I_{cd}$, when $m_{h}\gg m_{k}$, are found by
considering the terms with coefficient $m_{h}^{2}$ in
(\ref{i_cd_vanderbij_expressions}). At higher order in
$(m_{k}/m_{h})^{2}$ contributions from the terms with coefficient
$m_{k}^{2}$ will also become important. Note that
(\ref{leading_i_large_h}) is independent of $m_{k}$ and the lepton
masses and that its form may be understood as follows. Define:
\begin{eqnarray}
I_{2}(k^{2})&\equiv& \frac{-k^{2}}{\pi^{4}}\int d^{4}p\int
d^{4}q\frac{1}{(p^{2}-m^{2})}\frac{1}{(p-k)^{2}}\nonumber\\
& &\mkern40mu\times\frac{1}{(q^{2}-m^{2})}\frac{1}{(k-q)^{2}}\frac{1}{(p-q)^{2}}.
\end{eqnarray}
This corresponds, up to multiplicative factors, to $I_{cd}$ with
\mbox{$m_{c}=m_{d}=m_{k}=0$} and \mbox{$m_{h}\rightarrow m$} when the
external momenta $k$ is taken to be zero:
\begin{eqnarray}
\left.\left.I_{cd}\right|_{m_{c}=m_{d}=m_{k}=0} =\lim_{k^{2}\rightarrow
  0}\frac{1}{(2\pi )^{8}}\left( -\frac{\pi^{4}}{k^{2}}\right)
  I_{2}(k^{2})\right|_{m\rightarrow m_{h}}\nonumber.
\end{eqnarray}
The integral $I_{2}(k^{2})$ has been
evaluated~\cite{broadhurst}, leading to the result:
\begin{eqnarray}
\left. I_{cd}\right|_{m_{c}=m_{d}=m_{k}=0}
=\frac{1}{(4\pi)^{4}}\frac{2}{m^{2}_{h}}\zeta (2). \label{i_mc_md_mk_zero}
\end{eqnarray}

Noting that \mbox{$\zeta (2)=\frac{\pi^{2}}{6}$} and comparing
(\ref{i_mc_md_mk_zero}) with (\ref{leading_i_large_h}) it is seen that
the leading term of $I_{cd}$, for \mbox{$m_{h}\gg m_{k}$}, is simply the
result one obtains for $I_{cd}$ when the lepton masses and $m_{k}$ are
taken to be zero (observe that for $m_{h}\gg m_{k}$ we may safely set
$m_{c}$, $m_{d}$ and $m_{k}$ to zero to obtain the leading term as
$I_{cd}$ remains infra-red finite in this limit).

After completing this work we became aware that the explicit
asymptotic terms in
equations (\ref{i_cd_large_k}) and (\ref{leading_i_large_h}) were
given in~\cite{babu_macesanu}.
\section{Conclusion}
The analytic result for the integral:
\begin{eqnarray}
I_{cd} &=& \int \frac{d^{4}p}{(2\pi)^{4}}\int
\frac{d^{4}q}{(2\pi)^{4}}\frac{1}{(p^{2}-m_{h}^{2})}\frac{1}{(p^{2}-m_{c}^{2})}\frac{1}{(q^{2}-m_{h}^{2})}\nonumber\\
& &\times\frac{1}{(q^{2}-m_{d}^{2})}\frac{1}{(p-q)^{2}-m_{k}^{2}},
\end{eqnarray}
which appears in the lowest order neutrino mass matrix in Babu's
model, has been obtained. The result comprises of equation
(\ref{i_cd_vanderbij_expressions}) together with (\ref{mm_mone_mtwo}) and either (\ref{f_ab}) or
(\ref{my_form_of_f}) for the terms $(2m|m_{1}|m_{2})$. We note that
for $m_{k}\gg m_{h}$ the leading term is:
\begin{eqnarray*}
I_{cd}\approx
\frac{1}{(4\pi)^{4}}\frac{1}{m_{k}^{2}}\log^{2}\left(\frac{m_{h}^{2}}{m_{k}^{2}}\right).
\end{eqnarray*}
The
$m_{k}^{-2}$ term in the asymptotic expansion was extracted and is
in agreement with the result of~\cite{babu_macesanu}. The leading
term when $m_{k}\gg m_{h}$ is independent of the lepton masses but requires a non-zero value
of $m_{h}$ to avoid an infra-red singularity. 
For the reversed hierarchy, $m_{h}\gg m_{k}$, the leading term is:
\begin{eqnarray*}
I_{cd}\approx\frac{1}{(4\pi)^{4}}\frac{2}{m_{h}^{2}}\frac{\pi^{2}}{6},
\end{eqnarray*}
which is the result one obtains for $I_{cd}$ upon setting the lepton
masses and $m_{k}$ to zero and is in agreement with that obtained in~\cite{babu_macesanu}.
\section*{Acknowledgements}
This work was supported in part by the Australian Research Council. We
thank Dr. Macesanu for drawing our attention to reference~\cite{babu_macesanu}.


\begin{thebibliography}{999}
\bibitem{zee} A. Zee, Phys. Lett. 
\mbox{$\mathbf{B93}$}, 389 (1980);
\bibitem{wolf_zee} L. Wolfenstein, Nucl. Phys. 
\mbox{$\mathbf{B175}$}, 93 (1980);
\bibitem{babu} K.S. Babu, Phys. Lett.
\mbox{$\mathbf{B203}$}, 132 (1987);
\bibitem{petkov_toshev} S.T. Petcov and S.T. Toshev, Phys. Lett.
\mbox{$\mathbf{B143}$}, 175 (1984);
\bibitem{babu_ma} K.S. Babu and E. Ma, Phys. Rev. Lett.
\mbox{$\mathbf{61}$}, 674 (1988);
\bibitem{gracey} D. Choudhury,
R. Gandhi, J.A. Gracey and B. Mukhopadhyaya, Phys. Rev.
\mbox{$\mathbf{D50}$}, 3468 (1994);
\bibitem{vanderbij_veltman} J. van der Bij and M. Veltman, Nucl. Phys.
\mbox{$\mathbf{B231}$}, 205 (1983);
\bibitem{lahanes_tamvakis_vayonakis} A.B. Lahanes, K. Tamvakis and
  C.E. Vayonakis, Nucl. Phys. 
\mbox{$\mathbf{B196}$}, 11 (1982);
\bibitem{broadhurst} D.J. Broadhurst, Z. Phys. C - Particles and Fields 
\mbox{$\mathbf{47}$}, 115 (1990);
\bibitem{davydychev_tausk} A.I. Davydychev and J.B. Tausk,
  Nucl. Phys. 
\mbox{$\mathbf{B397}$}, 123 (1993);
\bibitem{ford_jones} C. Ford and D.R.T. Jones, Phys. Lett.
\mbox{$\mathbf{B274}$}, 409 (1992);
\bibitem{ford_jack_jones} C. Ford, I. Jack and D.R.T. Jones, Nucl. Phys. 
\mbox{$\mathbf{B387}$}, 373 (1992);
\bibitem{'thooft_veltman} G. 't~Hooft and M. Veltman, Nucl. Phys. 
\mbox{$\mathbf{B44}$}, 189 (1972); 
\bibitem{ghinculov_vanderbij} A. Ghinculov and J.J. van der Bij, Nucl. Phys. 
\mbox{$\mathbf{B436}$}, 30 (1995);
\bibitem{coleman_norton} S. Coleman and R.E. Norton, Nuovo Cimento 
\mbox{$\mathbf{38}$}, 438 (1965);
\bibitem{babu_macesanu} K.S. Babu and C. Macesanu, Phys. Rev.
\mbox{$\mathbf{D67}$}, 073010 (2003);
\end{thebibliography}
\end{document}